\documentclass[prl,aps,a4,epsf,8pt,twocolumn,showpacs]{revtex4}

\usepackage{amsmath}
\usepackage{amssymb}
\usepackage{graphicx}
\begin{document}

\newcommand{\beq}{\begin{equation}}
\newcommand{\eeq}{\end{equation}}
\newcommand{\beqa}{\begin{eqnarray}}
\newcommand{\eeqa}{\end{eqnarray}}
\newcommand{\bmat}{\begin{displaymath}}
\newcommand{\emat}{\end{displaymath}}

\newcommand{\eq}[1]{Eq.~(\ref{#1})}

\newcommand{\lan}{\langle}
\newcommand{\ran}{\rangle}

\newcommand{\tav}[1]{\left\lan #1 \right\ran}
\newcommand{\sav}[1]{\left [\hspace*{-.13cm}\left| #1 \right|\hspace*{-.13cm}\right ]}
\newcommand{\lsav}{\Biggl [\hspace*{-.13cm}\Biggl | }
\newcommand{\rsav}{\Biggr |\hspace*{-.13cm}\Biggr ]_{\rm av} }

\title{Vortex stripe glass with self-generated randomness}

\author{Hajime Yoshino$^{1}$, Tomoaki Nogawa$^{2}$, Bongsoo Kim$^{3}$
}
\address{$^1$Department of Earth and Space Science, Faculty of Science,
 Osaka University, Toyonaka 560-0043, Japan\\
$^2$Department of Applied Physics. School of Engineering, 
The University of Tokyo, 7-3-1 Hongo, Bunkyo-ku, Tokyo 113-8656, Japan\\
$^3$Department of Physics, Changwon National University, 
Changwon 641-773, Korea.
}

\begin{abstract}
We found a finite temperature glass transition in the absence of quenched disorder in frustrated Josephson junction arrays (JJA) on 
a square lattice under magnetic field with {\it anisotropic} Josephson couplings by numerical simulations. The externally induced vortexes develop zigzag stripes into the direction of weaker coupling at low temperatures. The whole amorphous vortex solid can flow smoothly into the direction of stronger coupling by non-trivial soft modes. 
As the result the macroscopic phase coherence is destroyed being reminiscent of vortex flow in pure bulk superconductors or spin-chirality decoupling in frustrated magnets. On the contrary such soft-modes are absent into the direction of weaker coupling. 
Consequently the system behaves as a Ohmic (unjammed) liquid or superconducting (jammed) solid with respect to injection of 
external electric current, i.~e. shear, along different directions. 
This jamming-glass transition can be regarded as a two dimensional realization of the Aubry's transition.
\end{abstract}


\date{\today}
\maketitle

Glass transition is observed in diverse systems. 
An interesting basic question which continues to attract physicists is whether
{\it frustration} of either energetic or kinetic origins, which are considered to be indispensable to avoid crystallisation, is {\it sufficient} to 
realize formation of amorphous solids \cite{Sadoc,Tarjus-review,stripe-glass}. 
The so called irrationally frustrated Josephson junction array (JJA) on a square lattice with irrational number density $f$ of externally induced vortexes is an ideal system to clarify this view \cite{Halsey} since it possesses strong energetic frustration in a simplest form. While JJAs with rational $f$ develop periodic vortex lattices \cite{Teitel-Jayaprakash,ground-states}, such simple orderings may be avoided for irrational $f$. Indeed equilibrium relaxations were similar to the primary relaxations observed in typical fragile supercooled liquids \cite{JJA-relaxation}.

Another remarkable feature of the JJA is that its electric transport properties have formal analogy with the tribology and rheology : the current-voltage (IV) curves correspond to the flow curves, the shear-stress vs the shear-rate \cite{Yoshino}. In a recent work \cite{vortex-jamming} we observed that variation of {\it anisotropy} $\lambda$ of the Josephson coupling into $x$ and $y$-directions on the square lattice is analogous to the variation of the strength of interactions between incommensurate surfaces in friction models \cite{aubry,matsukawa-fukuyama} that exhibit the so called Aubry's transition which is a jamming-transition at zero temperature.
Such anisotropic JJAs of various $\lambda$  have already been fabricated by lithography techniques \cite{anisotropic-JJA}. 
In \cite{vortex-jamming} we have shown that the IV curves at $T=0$ with various $\lambda$ exhibit dynamical scaling features around the symmetric point $\lambda=1$ very similarly to the flow curves of granular systems \cite{granular-rheology} around the so called jamming point (point-J) \cite{Nagel-group}. 

In this letter, we present numerical evidences that the irrationally frustrated anisotropic JJA system exhibit a 2nd order phase transition into a glassy phase at a finite temperature $T_{c}(\lambda)$. 
Since there is no quenched disorder in the present system, the glass transition here is fundamentally different from those in the 
spin-glasses and conventional vortex glasses \cite{FFH} but closer to structural glass transitions of liquids. $T_{c}(\lambda)$ rapidly decreases with the decrease of the anisotropy $|\lambda-1|$ and quite possibly terminates at the 'point-J' $(T=0, \lambda=1)$, being consistent with recent studies \cite{Park-Choi-Kim-Jeon-Chung,Granato} which suggest the isotropic system with $\lambda=1$ remains in the vortex liquid state at any finite temperatures $T$.

\begin{figure}[t]
\includegraphics[width=0.45\textwidth]{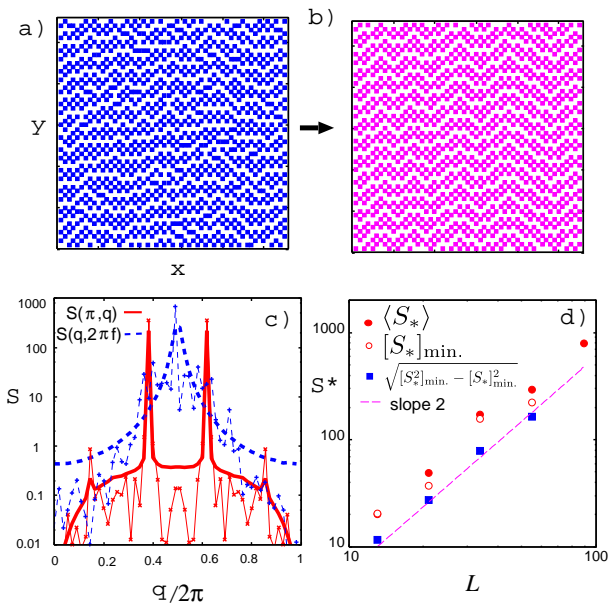}
\caption{Vortex patterns and the structure factor. 
Here $\lambda=1.5$ and $T=0.2$ ($T_{c} \sim 0.23$).
a) display the vortex pattern in a snapshot in equilibrium and b) displays that at a nearby energy minimum reached via an energy descent algorithm.
c) displays the cross-sections of the structure factor $S(q_{x},q_{y})$ ($L=55$) with thermal average (thick lines) and at the energy minimum shown in a) (thin lines).
d) displays the amplitude of the peak of the structure factor $S_{*}=S(\pi,2\pi f)$ with thermal average $\langle S_{*}\rangle$, average over the energy minima $[S_{*}]_{\rm min.}$ and variance of the minimum-to-minimum fluctuation $\sqrt{[S_{*}^{2}]_{\rm min.}-[S_{*}]_{\rm min.}^{2}}$. Here $[\ldots]_{\rm min}$ is took over $100$ energy-minima obtained by independent initial conditions.
} 
\end{figure}

{\bf Model} The JJA \cite{Tinkam,Teitel-Jayaprakash,Mooij-group} 
is a network of superconducting islands. We consider a square lattice of size $N=L^{2}$  and label the islands as $i=1,\ldots,N$.
The phase of superconducting order parameter $\theta_{i}$ of the $i$-th island 
is coupled to its nearest neighbors by the Josephson junctions.
External transverse magnetic field $B$ threads the cells in the forms 
of flux lines each of which carrying a flux quantum $\phi_{0}=c h/2e$.
On average each unit cell with area $a^{2}$ carries $f= B a^{2}/\phi_{0}$ flux lines. A flux line threading a unit cell induces a vortex of the phases 
around the cell much as a dislocation in a crystal. 
The number density $f$ is chosen to be an irrational number.
Thus the system is extremely frustrated such that even formation
of periodic vortex (dislocation) lattice can be avoided.
The properties of the JJA under the transverse magnetic field
are known to be described by an effective Hamiltonian \cite{Tinkam,Teitel-Jayaprakash},
\beq
H= -\sum_{<i,j> \parallel x\mbox{-axis}} \cos(\psi_{ij})
 - \lambda \sum_{<i,j> \parallel y\mbox{-axis}} \cos(\psi_{ij})
\label{eq-jja}
\eeq
with the gauge-invariant phase difference, $\psi_{ij} \equiv \theta_{i}-\theta_{j}-A_{ij}$.
We measure temperature $T$ in a unit with $k_{\rm B}=1$.
Here $\lambda$ is the ratio of the strength of the Josephson couplings along $x$ and $y$-axis. 
In the following we only examine $\lambda \geq 1$, which is obviously sufficient by symmetry. 
The vector potential $A_{ij}(=-A_{ji})$  is defined such that directed sum of them around each cell is  $2\pi f$.

Charge $v_{i}$ of the vortex  at the cell associated with the $i$-th lattice site is defined by taking directed sum of $(\psi_{ij}-[\psi_{ij}]_{n})/2\pi$ on the junctions around the cell. Here $[x]_{n}$ denotes the nearest integer of the real variable $x$. It takes values $\ldots,-1+f,f,1+f,\ldots$. The total charge is zero due to the charge neutrality.

{\bf Methods} In numerical simulations, we used a series of rational numbers $p/q=5/13$, $8/21$, $13/34$, $21/55$, $34/89$, $55/144$, $89/233$ for the filling $f$,  which approximate an irrational number $f=(3-\sqrt{5})/2$=0.38196601... We worked on systems of size $L \times L$ with $L=q$ so that in the thermodynamic limit $L \to \infty$ the value of $f$ converges to the target irrational number. 

To investigate macroscopic static properties, we used Monte Carlo (MC) simulations. We performed simulations on $L=13-89$ using $20-120$ temperatures in the temperature range $T=0.2-0.4$ by the exchange MC method \cite{Hukushima-Nemoto}. More precisely each MC step consists of one sweep over the system by the Metropolis updates of each phase $\theta_{i}$, one step of the overrelaxation \cite{Creutz} and one step of the exchange MC. We used $10^{5}-10^{6}$ MC steps for the equilibration and observations.

We also measured two-time auto-correlation functions during relaxation 
from random initial configurations by a simple Metropolis updates
over $10^{6}$ MC steps on larger systems with $L=144,233$ which do not 
show appreciable finite size effects within the simulation time scale.

\begin{figure}[t]
\includegraphics[width=0.45\textwidth]{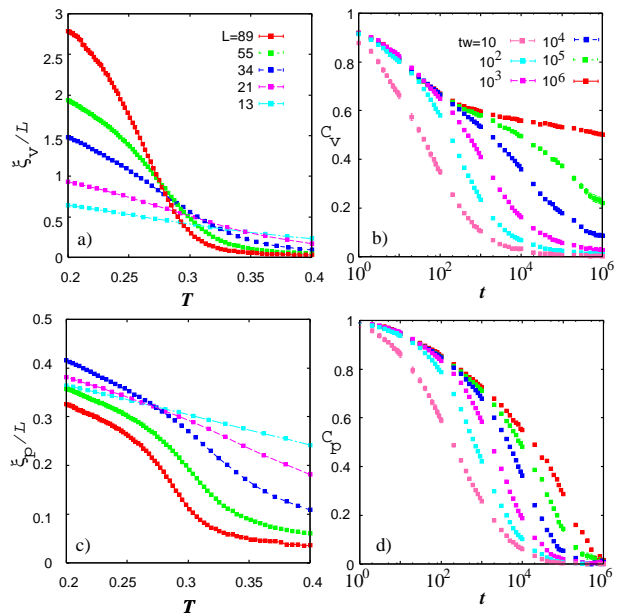}
\caption{Correlation lengths and two-time auto-correlation functions.
a) Correlation length of the vortex $\xi_{\rm V}$ and c) phase $\xi_{\rm P}$. 
Here $\lambda=1.75$ for which $T_{c} \sim 0.25$. b) Auto-correlation functions of vortex $C_{\rm V}(t+t_{\rm w},t_{\rm w})$ and d) phase $C_{\rm P}(t+t_{\rm w},t_{\rm w})$ during aging at $T=0.2$ after temperature quench with waiting times $t_{\rm w}=10$,$10^{2}$,$10^{3}$,$10^{4}$,$10^{5}$ and $10^{6}$ from left to right curves.} 
\end{figure}

{\bf Vortex stripe glass}
As shown in Fig.~1 a), the vortexes develop zigzag folded layers along the direction of weaker coupling which are stacked along the direction of stronger coupling at low temperatures. The formation of the smectic like stripes is reasonable because the repulsive interactions between vortexes are anisotropic. A remarkable feature is that the zigzag folds at different layers are strongly correlated. Indeed, as shown in Fig.~1 b), the alignment of the zigzag pattern is nearly perfect in the nearby energy minimum which we obtained by a simple energy descent algorithm. As shown in  Fig.~1 c) d), the structure factor $S(q_{x},q_{y})$ exhibits prominent peaks at $(q_{x},q_{y})=(\pi,2\pi f)$ and $(\pi,2\pi(1-f))$ whose height scales with the system size as $N=L^{2}$ as usual Bragg peaks. However the profile of the peak is peculiar: it decays sharply along $q_{y}$ but decays slowly by a power law $|\pi-q_{x}|^{-2}$ along $q_{x}$. The exponent is essentially independent of the temperatures. These features reflect the random zigzag wandering of layers which are strongly correlated between different layers. Somewhat different smectic layering of vortexes exist in bulk high-Tc superconductors due to the presence of the oxide layers \cite{smectic-vortex-highTc}.

Fig.~2 a) displays the scaled correlation length $\xi_{V}/L$ of the vortexes measured by the same method as in \cite{Granato}. (See \cite{Granato} for the details of the method, which is a standard technique for spin-glasses.) Here the correlation length along the direction of stronger coupling is shown. The scaled correlation length $\xi_{\rm V}/L$ of different sizes exhibit crossings suggesting a 2nd order phase transition at $T_{c}(\lambda)$, at marked variance with the isotropic $\lambda=1$ case \cite{Granato} in which $\xi_{\rm V} \sim T^{-2.2}$.

{\bf Absence of phase long-ranged order}
For the case of isotropic coupling ($\lambda=1$), it has been noticed that correlation length of the phase grows less rapidly  $\xi_{\rm P} \sim T^{-1.2}$ than that of the vortex $\xi_{\rm V} \sim T^{-2.2}$. This is very similar to the spin-chirality decoupling observed in a 2 dimensional XY spin-glass \cite{Kawamura-Tanemura} in the sense that chirality is equivalent to the externally induced vortex.

We found the decoupling is much more enhanced in the presence of anisotropy.
 Fig.~2 b) displays the scaled correlation length $\xi_{\rm P}/L$ along the direction of 
stronger coupling. The data do not exhibit crossings at larger $L$.
Furthermore the two-time phase auto-correlation function 
$C_{\rm P}(t+t_{\rm w},t_{\rm w})=(1/N)\sum_{i=1}^{N}\langle 
\cos\big(\theta_{i}(t+t_{\rm w})-\theta_{i}(t_{\rm w})\big)\rangle$ do not exhibit any plateau as shown in Fig.~2 d).
These features suggest absence of phase long-ranged order.

{\bf Ergodicity breaking} Starting from different initial conditions we could obtain
numerous energy minima similar to the one shown in Fig.~1 b) but with different random configurations of the zigzag fold. The randomness is reflected for example in the minimum-to-minimum fluctuations of the structure factor shown in Fig.~1 c) d). Note that both the average and the variance of the fluctuation grows linearly with the system size $N$.Quite remarkably we found the energy differences among them are only $O(1)$ so that all of them are relevant in the static equilibrium ensemble. However, the nearly perfect vertical alignment of the zigzag fold suggests that it would take extremely long time to change the pattern of the zigzag fold from one to another dynamically. 

To clarify this point, we performed aging simulations and measured the two-time vortex auto-correlation function $C_{\rm V}(t+t_{\rm w},t_{\rm w})=(1/N)\sum_{i=1}^{N}\langle v_{i}(t+t_{\rm w})v_{i}(t)\rangle$.  As shown in Fig.~2 b) it exhibits a plateau at larger waiting times $t_{\rm w}$, which suggests the Edwards-Anderson order parameter of the vortex $q_{\rm V} \equiv \lim_{t \to \infty}\lim_{t_{\rm w} \to \infty} C_{\rm V}(t+t_{\rm w},t_{\rm w})$ is finite, i.~e. the ergodicity is broken. It implies that once the system reaches an energy minimum, as the one shown in Fig.~1, it cannot escape out of it. Thus in sharp contrast to the usual smectic liquid crystals \cite{Chaikin-Lubensky}, the transverse wandering of the layers are frozen in time. Note that usual smectic liquid crystals have only quasi-long raged orders along all directions so that the layers are not so regularly stacked as in the present system. 

\begin{figure}[t]
\includegraphics[width=0.45\textwidth]{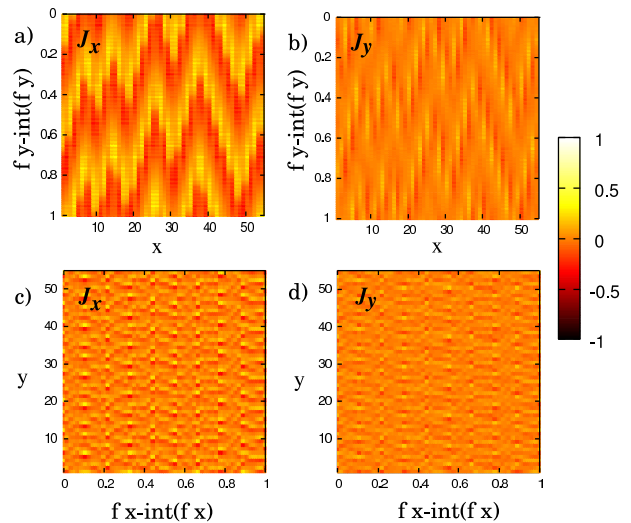}
\caption{Spatial configuration of currents through the Josephson junctions 
in an energy minimum (shown in Fig.~1 b)). 
Coordinate along either $x$ or $y$ axis is ``folded'' to 
elucidate modulation with respect to ideal incommensurate ordered states (see text).
Panels a) b) show Josephson currents into $x$ and  $y$ directions plotted against 
the folded $y$-axis. Similarly panels c) d) show the currents
against the folded $x$-axis. 
}
\end{figure}

{\bf Soft modes} To obtain some insights on the low lying metastable states, 
we analyzed the spatial configuration of the Josephson currents $\sin(\psi_{ij})$. 
Since the vortexes have long ranged order with the incommensurate wave number $q_{y}=2\pi f$ into the direction of stronger coupling, 
which is the $y$-direction for $\lambda >1$, we may take ideal incommensurate ordered states as references.
Then it is convenient to introduce a  ``folded coordinate'' $f y -{\rm Int} (f y)$ 
with ${\rm Int}(x)$ being a function which truncates a real number $x$ to an integer number 
\cite{note}.  

Fig.~3 a) b) display the configuration of the current $J_{x}$ and $J_{y}$ along 
the junctions parallel to the $x$ and $y$ axises. 
Apparently the currents behave as a continuous function of the folded $y$-axis. 
This immediately means existence of soft-modes: by choosing an arbitrary real number $\alpha$ 
and changing the entire configuration such that the value of the gauge invariant phase 
difference $\psi$  at $(x,f y -{\rm Int} (f y))$ is replaced by that at 
$(x,f (y+\alpha) -{\rm Int} (f (y+\alpha))$ we obtain a different configuration. 
By a little thought one can find that it is another energy minimum with exactly the same energy. 
We note that $J_{x}$ takes all possible values between $-1$ and $1$ while $J_{y}$ 
takes values only within a narrow range around $0$. 
It means that the shift by a certain $\alpha$ in the folded $y$-coordinate amount 
to put shear across the $x$-direction by which the whole vortex solid slide along 
the $y$-direction without changing its pattern and the energy: in spite of the fact that the vortexes live on the lattice, they flow like in the pure bulk superconductors \cite{Tinkam}.

In contrast, the configuration of the currents with the folded $x$-axis do not
reveal any continuous functions but discontinuous ones as shown in Fig.~3 c) d). 
This means the vortex solid cannot slide freely along the $x$-direction.

The presence/absence of the soft modes into different directions naturally
explain the unusual features observed in \cite{vortex-jamming} that shear (helicity) modulus is finite/zero with respect to shear  (injection of external electric current) along the direction of stronger/weaker coupling (See Fig.~4 b)). The isotropic point $\lambda=1$ appears as a critical point ('J-point') for the shear response at $T=0$. Correspondingly dynamic shear response also exhibit jammed/unjammed behaviours as shown in \cite{vortex-jamming}. 

Furthermore the phase coherence of the system along the direction of weaker coupling 
must be destroyed by the soft mode by which the vortex solid slide into the direction 
of the stronger coupling. 
Thus the phase coherence can only grow in one dimensional way along the direction of 
stronger coupling resulting in the absence of long-ranged phase order, as observed here.

\begin{figure}[t]
\includegraphics[width=0.45\textwidth]{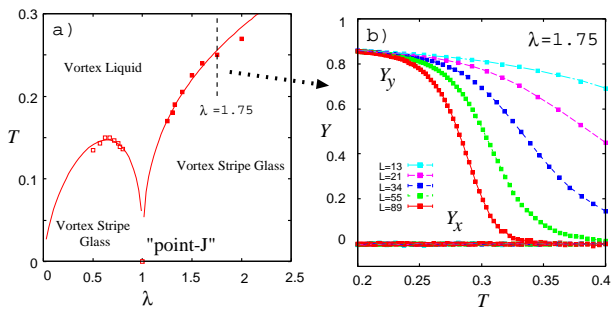}
\caption{Phase diagram and helicity modulus.
a) Note that $T_{c}(1/\lambda)=T_{c}(\lambda)/\lambda$ holds
due to the obvious symmetry between $x$ and $y$ axis. 
b) The profile of helicity modulus $Y_{x}$ and $Y_{y}$ for shear along
$x$ and $y$ directions at $\lambda=1.75$.
}
\end{figure} 

{\bf Aubry's transition as a jamming-glass transition}
In an one dimensional limit the 2 dimensional JJA reduces to a JJA on a two-leg ladder lattice \cite{Denniston-Tang-ladder}, which can be viewed as an Eulerian representation of models
for friction between two atomic layers \cite{aubry,matsukawa-fukuyama}. 
The ratio of the atomic spacings of the layers is assumed to be an irrational number $f$, which is identified to the vortex density $f$ in the JJA after some gauge transformations. 

The Josephson currents $\sin(\phi_{ij})$ shown in Fig,~3 correspond to inter/intra layer forces between atoms in the friction models. 
In the Frenkel-Kontorova model it is known rigorously that they are analytic functions of the folded coordinate in the unjammed phase but non-analytic functions with dense set of discontinuous points in the jammed phase where the static frictional force or yield stress (corresponding to the critical current in JJA) is finite \cite{aubry}. They are called as the Hull functions. 

Our results discussed above strongly suggest the jamming-glass transition we found in the present model is a two-dimensional analogue of the Aubry's transition. In contrast to the one-dimensional models, however, the glassy phase is observed not only at zero temperature but also at finite temperatures. 
In Fig.~4 a) we display the phase diagram of the present system. 
The critical temperature $T_{c}(\lambda)$ is estimated by measurement of 
the relaxation times $\tau_{\rm V}(T,\lambda)$ of the vortex auto-correlation 
function $C_{\rm V}(t)$ which 
can be fitted as $\tau_{\rm V}(T,\lambda) \sim (T-T_{\rm c}(\lambda))^{-z\nu}$ 
with $z\nu \sim 6.5$ and
$T_{c}(\lambda) \sim |\lambda-1|^{\rho}$ with $\rho \sim 0.4$. 
More details of the results will be reported elsewhere.

\vspace*{.2cm}
{\bf Acknowledgement} We thank Hikaru Kawamura and Hiroshi Matsukawa for
useful discussions.
The authors thank the Supercomputer Center, Institute for Solid State Physics, 
University of Tokyo for the use of the facilities.
This work is supported by Grant-in-Aid for Scientific Research
on Priority Areas "Novel States of Matter Induced by Frustration"
(1905200*) and Grant-in-Aid for Scientific Research (C) (21540386).


\begin{thebibliography}{99}
\bibitem{note} All coordinate points on the original $y$-coordinate fall into the range $[0,1)$ of the folded $y$-coordinate. Since $f$ is an irrational number, the original discrete coordinate points uniformly fill the entire range $[0,1)$ in the thermodynamic limit $L \to \infty$, which we assume below for simplicity.
\end{thebibliography}
\end{document}